\documentclass[english]{lni}

\IfFileExists{latin1.sty}{\usepackage{latin1}}{\usepackage{isolatin1}}

\usepackage{amsmath,graphicx,multirow,booktabs,color,colortbl}
\usepackage{layout}
\usepackage{fancyhdr}
\usepackage{listings} 
\usepackage{changepage} 
\usepackage[figurename=Fig., tablename=Tab., small]{caption}[2008/04/01]

\renewcommand{\headrulewidth}{0.4pt} 
\author{Per B{\ae}kgaard\footnote{Cognitive Systems, DTU Compute; Technical University of Denmark, DK-2800 Kgs. Lyngby; pgba@dtu.dk}\, Michael Kai Petersen\footnote{Cognitive Systems, DTU Compute; Technical University of Denmark, DK-2800 Kgs. Lyngby; mkai@dtu.dk} \, Jakob Eg Larsen\footnote{Cognitive Systems, DTU Compute; Technical University of Denmark, DK-2800 Kgs. Lyngby; jaeg@dtu.dk} }

\title{The Blank Stare: Retrieving Unique Eye Tracking Signatures Independent of Visual Stimuli}

\begin{document}


\maketitle

\renewcommand{\refname}{References}
\setcounter{footnote}{3} 

\begin{abstract}
\textsc{Using Low Cost Portable Eye Tracking for Biometric Identification or Verification:} 
Eye tracking technologies have in recent years become available outside of specialised labs, and are starting to become integrated in tablets and virtual reality headsets\footnote{E.g. at http://www.theeyetribe.com}. This offers new opportunities for use in common office- and home environments, such as for biometric recognition (identification or verification), alone or in combination with other technologies. This paper exposes two fundamentally different approaches that have been suggested, based on spatial and temporal signatures respectively. While deploying different stimulation paradigms for recording, it also proposes an alternative way to analyze spatial domain signatures using Fourier transformation. Empirical data recorded from two subjects over two weeks, three months apart, are found to support previous results. Further, variations and stability of some of the proposed signatures are analyzed over the extended timeframe and under slightly varying conditions.
\end{abstract}



\section{Introduction}\label{introduction}

Biometric identification and verification are applied and subject to active standardization work \cite{iso19794}; in particular within the areas of fingerprints (minutiae) and face- and iris images, and is extending to dynamic signature time series, vascular images, hand geometry silhouette, voice data, DNA data and palm crease images. 
Less work has seemingly been done on the inherently traceless, non-invasive and contactless \textit{eye tracking signatures} like \textit{fixations} and \textit{saccades} \cite{holmqvist2011eye} for behavioural biometric purposes.

\textbf{Temporal Features -- Oculomotor Plant Model Analysis of Saccades:}\label{oculomotor-plant-model-analysis-of-saccades}
\cite{komogortsev2010biometric}, \cite{komogortsev2012biometric} and \cite{komogortsev2014biometrics} however suggest methods for biometric identification and verification using eye trace recordings. These are largely based on observing eye movement traces from which individual \textit{saccades} are used to estimate parameters of a proposed \textit{Oculomotor Plant Model} (OPM), thus serving as an extracted feature vector from the biometric probe, comparable to previous feature vectors of one or more biometric references from a biometric enrolment database. In the more recent work \cite{komogortsev2014biometrics}, \textit{Equal Error Rates} (EER) of the best model was found at 20.3\% and \textit{Receiver Operating Characteristics} (ROC) \textit{Area Under Curve} (AUC) levels of up to around 80\% were observed in a study of 32 users performing a total of 122 unique recordings within a period of 2 weeks.

Some of the key advantages of using such a system\footnote{Other, although apparently few, proposals that also explore biometric features extraction from eye movements exists; see the introduction section of \cite{rigas2014biometric} for a summary of some of the key ideas.} are that it is difficult for a subversive impersonator in a presentation attack to spoof dynamic patterns that would result in a feature vector matching the intended reference; such a system will often include a liveness detection as salient on-screen stimuli can be used to trigger saccades that would be difficult for a non-human entity to mimic. Further, parameters of such a system could be estimated on a continuous basis, to ensure that a user carrying out trusted tasks is not replaced after verification. It could also be speculated that the collectability and acceptability might be high, due to the unobtrusive and readily available nature of the technology.

\pagestyle{fancy}
\fancyhead{} 
\fancyhead[RO]{\small The Blank Stare \hspace{5pt} \thepage \hspace{0.05cm}}
\fancyhead[LE]{\hspace{0.05cm}\small \thepage \hspace{5pt} Per B{\ae}kgaard et.al.}

\fancyfoot{} 
\renewcommand{\headrulewidth}{0.4pt} 

\textbf{Spatial Features -- Fixation Density Map Analysis:}\label{fixation-density-map-analysis}
The method(s) above work best when a large number of 
saccades can be evoked and analysed. 
Where the OPM based proposal(s) mainly rely on exciting, and estimating characteristics of, the dynamic eye-brain system by presenting salient stimuli, the work of \cite{rigas2014biometric} proposes the use of \textit{fixation density maps} (FDM) for biometric identification or verification in a way that does not rely on invoking specific eye movements.
It rather explores characteristics of the spatial domain,
which \cite{baekgaard2015thinking} also found surprisingly stable within two subjects during a week of repeated experiments.

\cite{rigas2014biometric} analyses the FDM that result from observing stimuli in a free-viewing condition, noting that the FDMs also embeds some temporal information as the density grows with the amount of time the eye remains within an area. The FDMs of 15 subjects observing faces, 100 subjects reading text and the same 100 subjects observing video-scenes were collected and analysed, and a min-based metric as well as Euclidean Kullback-Leibler divergence (KLD) \textit{dissimilarity scores} were used to compare the FDMs and to compute \textit{dissimilarity matrices}. The effect of applying different size gaussian kernels on the FDM were also evaluated. 
In the best possible configuration, EER rates of 18.3\% were reported, with variations up to 34.5\%, but still above chance level. 

\section{Biometric Brain-Eye Feature Extraction}\label{biometric-brain-eyefeature-extraction}

Extracted temporal- and spatial features complement each other well. Whereas the temporal features to a large degree reflect the dynamic responses of the brain-eye system with its neuronal-nerve-muscle interactions, the spatial features are, to a much larger degree, independent of many of the OPM parameters, and thus may reflect more of the processes related to components of the \textit{attention networks} in the brain.

The present work looks at both temporal- and spatial feature extraction and comparisons, using low-cost eye tracking equipment in office conditions where subjects are free to move around somewhat and/or be distracted temporarily by other events in the environment.
In addition to replicating selected aspects of previous works, it also hints at the longer-term stability of the analysed features, and proposes applying a \textit{Discrete Fourier Transformation} (DFT) to better highlight desired attention-driven qualities and find similarities in the spatial domain.

\textbf{Temporal (Bottom-Up) Features -- Saccades and Time-to-Target:}\label{temporal-bottom-up-features-and-saccades}
For the temporal domain, a simple feature extraction and analysis is done: From the OPM model, it can be expected that the effective \textit{time-to-target} (TTT) or \textit{saccadic latency}, measured from presentation of a salient stimuli until the first eye fixation has reached an area nearby, will depend on the direction of the eye movement and will vary between individuals. The hypothesis is that these TTT measures should remain relatively stable within subjects even over varying conditions and over longer periods.

\textbf{Spatial (Top-Down) Features -- Fixation Density Map Metrics:}\label{spatial-top-down-features-and-fixations}
For the spatial domain, this present work proposes a novel approach to analysis by applying a DFT to the FDM, which here appears to improve the comparison performance significantly.

Whereas \cite{rigas2014biometric} calculate spatial domain dissimilarity metrics directly from the resulting FDM after a gaussian kernel convolution, it is here proposed to first apply a DFT transformation, discarding the phase information and then use a box-filter to extract only the lower spatial-frequency components of the resulting spectrum before calculating feature vector distances. The filtering is not unlike applying a gaussian filter to the FDM, but the process as proposed discards translations of the FDM and focuses more on the spatial-frequency distributions instead of the actual FDM areas themselves, and may thus emphasize eye patterns, including some microsaccades, used when exploring areas of interest.

It is hypothesised that as the observed scene(s) present areas of interest in a given spatial configuration, the brain may explore different temporal and spatial combinations of these. The resulting FDM will likely not only differ between two users observing the same scene, but also to some extent when one user observe the same scene multiple times. However, if the top-down driven eye movements are more a result of individual preferences or characteristics related to attention governing the way a particular scene or areas of interest are explored, it might be possible to better extract and characterize some of these patterns in the DFT'ed domain.

The DFT function $F_{a,b}$ of a $N$x$N$ square 2D image, $f_{x,y}$, is:
$$
F_{a,b} = \sum_{x,y}^N f_{x,y} e^{-i2\pi ({{ax}\over{N}}+{{by}\over{N}})} 
$$
and applying a box-filter simply means assigning $F_{a,b}=0$ when $|a|>l \vee |b|>l$ for some limit $l$, chosen suitably to the size of the FDM (values around 5\% of the original image size has been used in the present case).

For comparison with \cite{rigas2014biometric}, a similar gaussian kernel is applied to the FDMs, but this is not strictly required for the DFT distance metrics to be calculated; only for the FDM domain distance metrics they are essential.

To compare the extracted 2D NxM features, 
consider a $P$ (probe) and a $R$ (reference) -- both based on either a FDM directly or a DFT'ed FDM -- for which the distance metrics are calculated. This will be done similarly to \cite{rigas2014biometric}, with the addition of a simple \textit{Mean Square Error} (MSE) metric.

Before calculating any of the distance metrics, it is ensured that the feature vectors are normed independently such that $\sum_{x,y} P_{x,y} = 1$ and $\sum_{x,y} R_{x,y} = 1$. The distance metrics are then as follows:

The \textit{Mean Square Error} (MSE) distance calculates the sum of the square of the difference at each 2D point in the feature space:
$$
D_{\operatorname{MSE}}(P,R)=\sum_{x,y} {(P_{x,y}-R_{x,y})}^{2} 
$$

The \textit{1-MIN based Similarity Metric} (1-Min) distance calculates the sum of the min value of either $P$ or $R$ at each 2D point in the feature space: 
$$
D_{\operatorname{1-MIN}} (P,R) = \sum_{x,y} \operatorname{min}(P_{x,y},R_{x,y}) 
$$
It is in this case, where the FDMs could be thought of as representing a probability function, somewhat similar to a continuous version of the Hamming distance that is conventionally deployed when only binary point-to-point comparisons are possible.

The \textit{Kullback-Leibler divergence} \cite{kullback1951information} (KLD), or relative entropy, is a non-symmetric measure defined as follows:
$$
D_{\operatorname{KL1}}(P,R)=\sum_{x,y} P_{x,y} ln {\frac {P_{x,y}}{R_{x,y}}}
$$
but in this present case, disregarding any performance considerations, a symmetric distance is desirable, and hence a value proportional to the harmonic mean of $D_{\operatorname{KL1}}(P,R)$ and $D_{\operatorname{KL1}}(R,P)$ is used, i.e. 
$$
D_{\operatorname{KLD}}(P,R) = \frac{2}{\frac{1}{D_{\operatorname{KL1}}(P,R)} + \frac{1}{D_{\operatorname{KL1}}(R,P)}}
$$

Finally, the \textit{Euclidean} (Eucl) distance is calculated, based on not only comparing two feature sets from FDMs or DFT'ed FDMs but using the entire previously calculated KLD dissimilarity matrix\footnote{
As the $D_{\operatorname{Eucl}}$ distance requires computing the entire dissimilarity matrix based on all participating feature vectors, it may be less suited to large biometric databases. It may work best with neither too small nor too large template databases against which a probe is to be compared. Also, the Euclidean distance between two feature vectors will change according to how many and which particular feature vectors are included in the total comparison, so a cut-off threshold learned from a training set cannot directly be applied to another test set; the classifications derived from the training set will have to be used to recalculate a new threshold value in the combined training and test sets.}
as follows: Since the $D_{\operatorname{KLD}}$ in our case is symmetric, the dissimilarity matrix will also be symmetrical. Each row (or equivalently column) can be considered as a feature vector that holds the $D_{\operatorname{KLD}}$ distance from the corresponding probe to all other templates\footnote{Fom this calculation, there is no difference as to whether compared samples are other probes or references.}. This can be compared to all other similar feature vectors by computing the Euclidean distance between each vector pair. Hence, a new dissimilarity matrix can be built by computing:
$$
D_{\operatorname{Eucl}} (D_{\operatorname{KLD}}) = \operatorname{Euclidean}(D_{\operatorname{KLD}}, D_{\operatorname{KLD}}^T)
$$
as proposed by \cite{rigas2014biometric}\footnote{In this case, $D_{\operatorname{KLD}}$ is symmetric and square, so the transposed matrix will be identical to the original matrix.} where it also is subsequently normed so that $max(D_{\operatorname{Eucl}}) = 1$

\section{Experimental Setup}\label{experimental-setup}

Two male right-handed subjects, average 55 years old, participated in the trials, which took place over several weeks\footnote{This footnote removed in order to keep the present work anonymous.}
\footnote{The first week of the experiment has been described and qualitatively analysed in \cite{baekgaard2015thinking}.} 
3 months apart\footnote{Although this does not prove much in terms of uniqueness or the overall permanence of the extracted features, it is nevertheless based on a longer period compared to previously noted work.}. In each week, the experiment was executed one or more times during most of the weekdays, at alternating hours and between two different everyday offices. In total 34 trials were run; 16 with subject A and 18 with subject B. The subjects were not instructed to follow any specific viewing patterns.

During the experiment 24 sequences were presented, where 8 colored squares (3 degrees wide) were sequentially presented as salient stimuli on the screen, each sequence alternating between the colors blue, yellow, green, yellow, white and black. Each presented square appeared for 2 seconds against their complementary color as background. This was followed by 4 seconds of solid complementary color\footnote{Hence the title of this paper \textsc{The Blank Stare}.}, in total 480 secs of visual stimuli for each of 34 experiments performed, in aggregate 2h16m of stimuli for each participant. 

The trial stimuli were presented on a conventional MacBook Pro 13” at 60 Hz screen refresh rate, running the PsychoPy software \cite{peirce2007}. The Eye Tribe mobile eye tracking device, connected via USB, was used to retrieve the eye traces through the associated API \cite{TheEyeTribeAPI}, using PeyeTribe \cite{PeyeTribe}, and was running at 60 Hz. An initial calibration, using the vendor provided interface, was performed at the beginning of each experiment. 

The raw traces from the trials were first analysed to identify fixations by applying a DBSCAN \cite{ester1996density} derived density-based clustering approach. They were then epoched according to the presence or lack of the stimuli squares, and FDMs were built from the fixations that occurred during the combined 4-second periods where only a solidly coloured background was present. This was used for the top-down FDM analysis.

For the bottom-up time-to-target analysis, the appearance or relocation of a stimuli square was used as the basis for calculating the time to the first fixation hereafter nearby the presented square. To remove outliers and be able to identify those occasions where the reaction could be assumed to be the result of a bottom-up response to the stimuli, time-to-target values of less than 0.1 sec or more than 0.4 secs were discarded as outliers\footnote{\cite{holmqvist2011eye} cites typical saccadic latencies of 200ms and never shorter than 100ms; the latter only found when attention is released before onset of a new stimuli (the \textit{gap effect}), which is not the case in this present study.}.

Apparent inaccurate calibrations, resulting in a shift of the fixation locations compared to the presented stimuli, are further compensated: A simple affine transformation can be calculated by minimizing the MSE from each measured vs target position, and can subsequently be applied to the fixations. These are the \textit{Recalibrated} scenarios, tested to see if they would result in better alignment and more comparable FDMs across trials.

\section{Results}\label{results}

\begin{figure}[htb]
\centering
\includegraphics[width=1.0\textwidth]{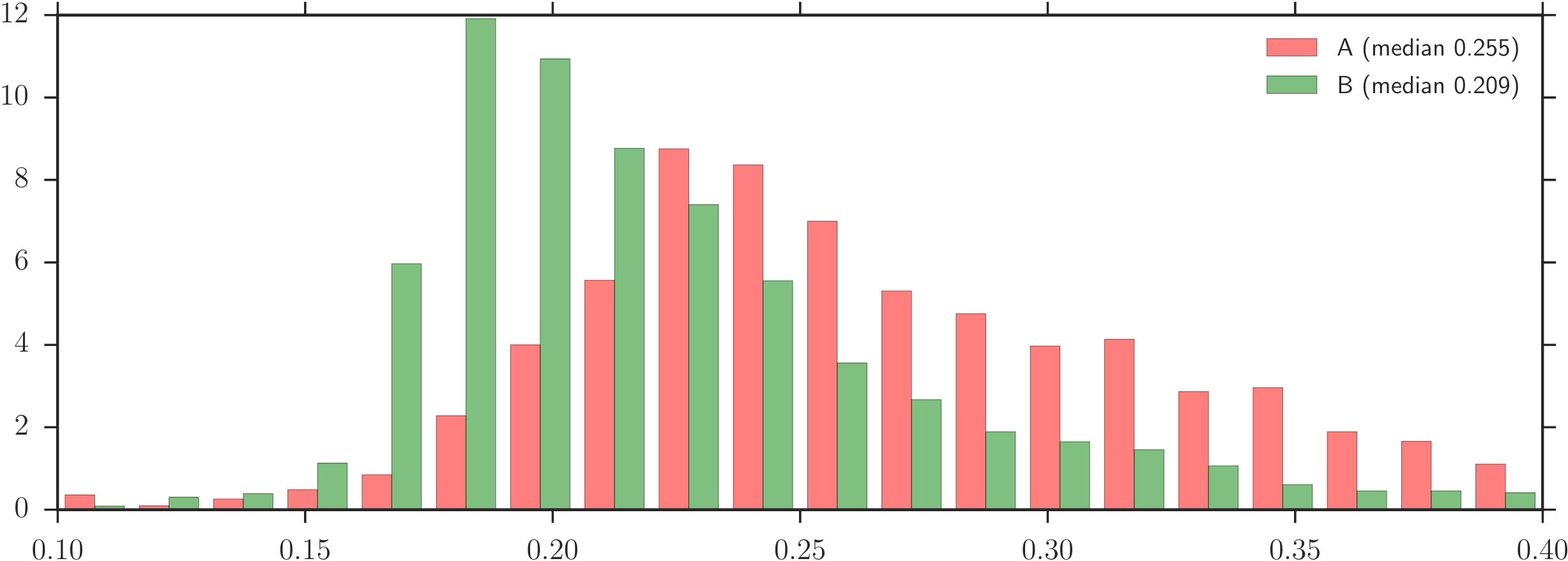}
\caption{Time-to-target histogram (all trials) for subject A (red) and subject B (green). The median value for subject A is 0.255s and for subject B 0.209s.}
\label{fig_rt_histo_all}
\end{figure}

\textbf{Temporal Features -- Time-to-Target:}\label{time-to-target}
Fig.~\ref{fig_rt_histo_all} shows the resulting median time-to-target values over all trials for both subjects. The distributions are different, but wide and with a long tail towards higher values. An individual time-to-target value cannot be contributed to either subject with confidence.

\begin{figure}[htb]
\centering
\includegraphics[width=1.0\textwidth]{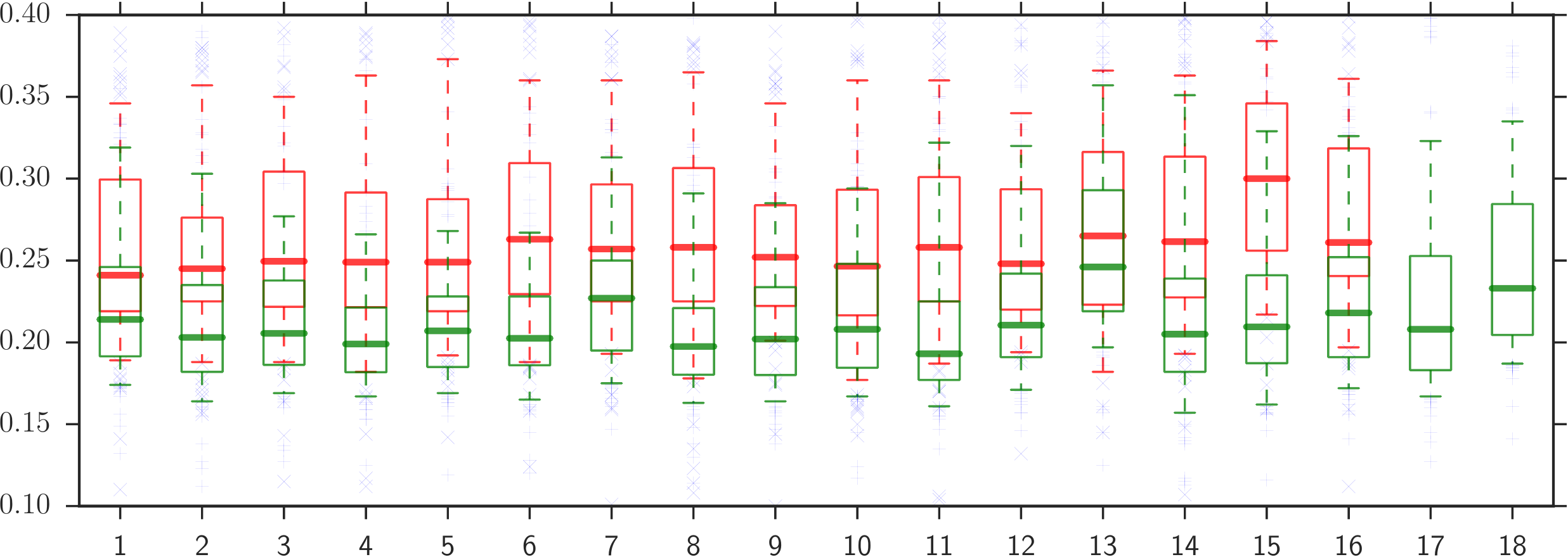}
\caption{Time-to-target for subjects A (red) and B (green). The bold line is the median, boxes outline quartiles, whiskers 5\% and 95\% confidence intervals. Numbers on the x-axis are for convenience.}
\label{fig_rt_overtrials-boxplot}
\end{figure}

Table \ref{tbl-ttt-all} show mean, median and standard deviation values for the time-to-target of all trials. As can be seen, even between trials, the distribution is quite wide and the differences between the mean or median values between subjects is of the same magnitude as the standard deviation. Hence, when comparing two trials, the difference is not statistically significant. The same data is also shown graphically in Fig.~\ref{fig_rt_overtrials-boxplot}.

\begin{figure}[htb]
\centering
\includegraphics[width=1.0\textwidth]{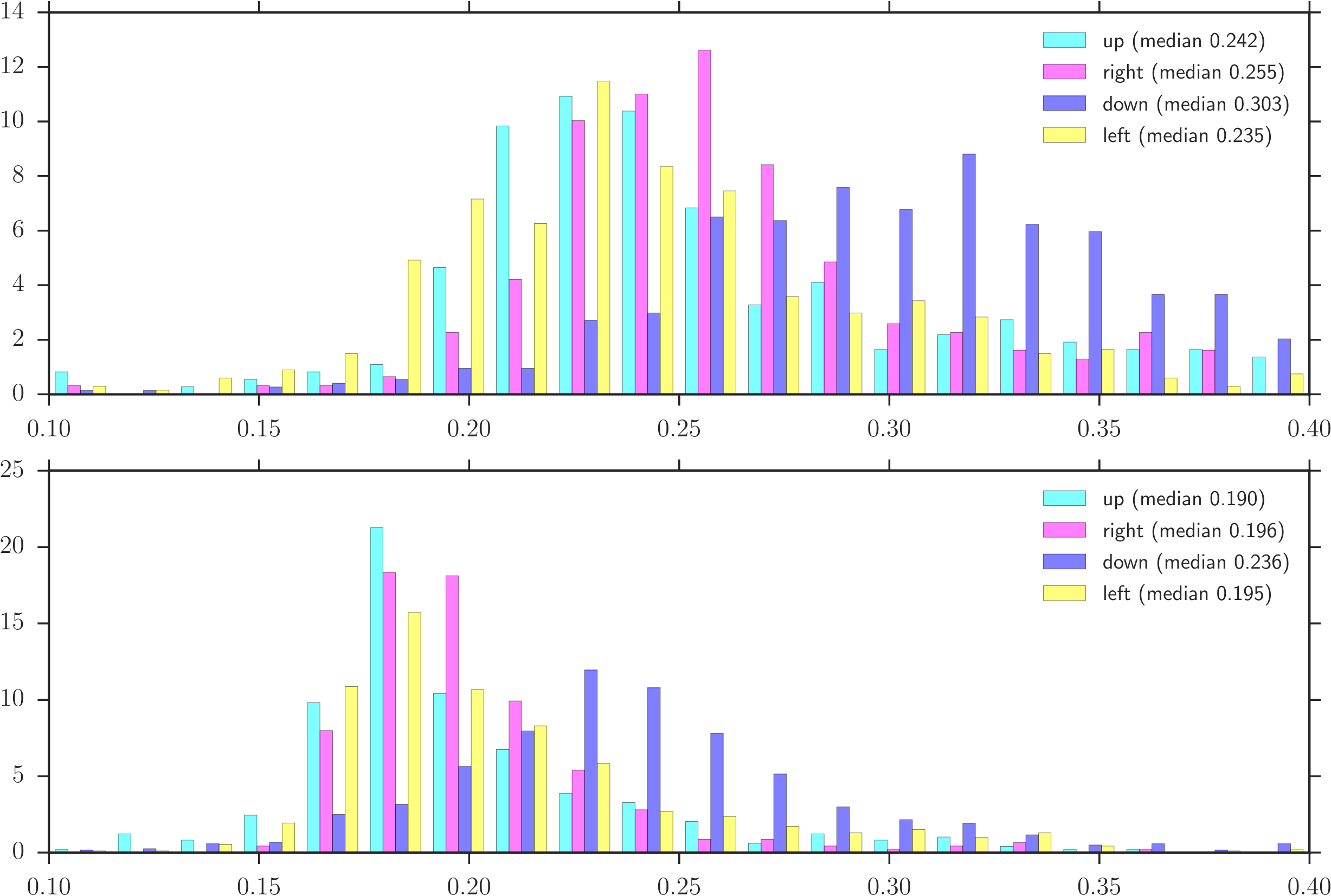}
\caption{Time-to-target histogram over all trials for subjects A \textsc{(Top)} and B \textsc{(Bottom)}. Cyan, magenta, blue and yellow bars correspond to upwards, right, downwards and left saccades, with medians 0.242s, 0.255, 0.303s and 0.235s for subject A and 0.190s, 0.196s, 0.236s and 0.195s for subject B.}
\label{fig_rt_histo_direction}
\end{figure}

Fig.~\ref{fig_rt_histo_direction} shows time-to-target when only looking at saccades in the upwards, right, downwards and left direction, respectively. The distributions are different within-subject when comparing the downward direction with the other three; for both subjects, downwards time-to-target values are somewhat longer (around 50ms) although the distributions still overlap. The other three time-to-target values are very similar within-subject.

\begin{table}[htb]
\centering
\resizebox{\textwidth}{!}{%
\begin{tabular}{@{}llrrrrrrrrrrrrrrrrrr@{}}
\toprule
 & Trial & \multicolumn{1}{c}{1} & \multicolumn{1}{c}{2} & \multicolumn{1}{c}{3} & \multicolumn{1}{c}{4} & \multicolumn{1}{c}{5} & \multicolumn{1}{c}{6} & \multicolumn{1}{c}{7} & \multicolumn{1}{c}{8} & \multicolumn{1}{c}{9} & \multicolumn{1}{c}{10} & \multicolumn{1}{c}{11} & \multicolumn{1}{c}{12} & \multicolumn{1}{c}{13} & \multicolumn{1}{c}{14} & \multicolumn{1}{c}{15} & \multicolumn{1}{c}{16} & \multicolumn{1}{c}{17} & \multicolumn{1}{c}{18} \\ \midrule
\multirow{4}{*}{A} & N & 151 & 152 & 112 & 119 & 131 & 147 & 139 & 143 & 154 & 148 & 153 & 75 & 60 & 152 & 93 & 119 &  &  \\
 & Mean & 0.255 & 0.255 & 0.261 & 0.256 & 0.260 & 0.270 & 0.264 & 0.263 & 0.258 & 0.256 & 0.265 & 0.256 & 0.269 & 0.270 & 0.298 & 0.276 &  &  \\
 & Median & 0.241 & 0.245 & 0.250 & 0.249 & 0.249 & 0.263 & 0.257 & 0.258 & 0.252 & 0.246 & 0.258 & 0.248 & 0.265 & 0.262 & 0.300 & 0.261 &  &  \\
 & $\sigma$ & 0.054 & 0.048 & 0.055 & 0.056 & 0.053 & 0.055 & 0.054 & 0.059 & 0.048 & 0.059 & 0.056 & 0.050 & 0.059 & 0.056 & 0.056 & 0.055 &  &  \\ \cmidrule(l){2-20}
\multirow{4}{*}{B} & N & 167 & 165 & 174 & 180 & 177 & 170 & 175 & 170 & 166 & 171 & 169 & 170 & 169 & 177 & 166 & 173 & 162 & 171 \\
 & Mean & 0.224 & 0.213 & 0.215 & 0.206 & 0.209 & 0.209 & 0.230 & 0.207 & 0.209 & 0.219 & 0.209 & 0.225 & 0.257 & 0.220 & 0.222 & 0.227 & 0.226 & 0.247 \\
 & Median & 0.214 & 0.203 & 0.205 & 0.199 & 0.207 & 0.202 & 0.227 & 0.198 & 0.202 & 0.208 & 0.193 & 0.210 & 0.246 & 0.205 & 0.209 & 0.218 & 0.208 & 0.233 \\
 & $\sigma$ & 0.045 & 0.046 & 0.042 & 0.036 & 0.032 & 0.037 & 0.045 & 0.042 & 0.040 & 0.043 & 0.049 & 0.049 & 0.053 & 0.059 & 0.052 & 0.046 & 0.058 & 0.051 \\ \cmidrule(l){2-20} 
\end{tabular}
}
\caption{Time-to-target values (all trials). N is the number of correctly identified target fixations within the 0.1s to 0.4s window. Mean, median and Standard Deviation are also listed.}
\label{tbl-ttt-all}
\end{table}

\begin{figure}[htb]
\centering
\begin{minipage}{0.45\textwidth}
\centering
\includegraphics[width=1.0\textwidth]{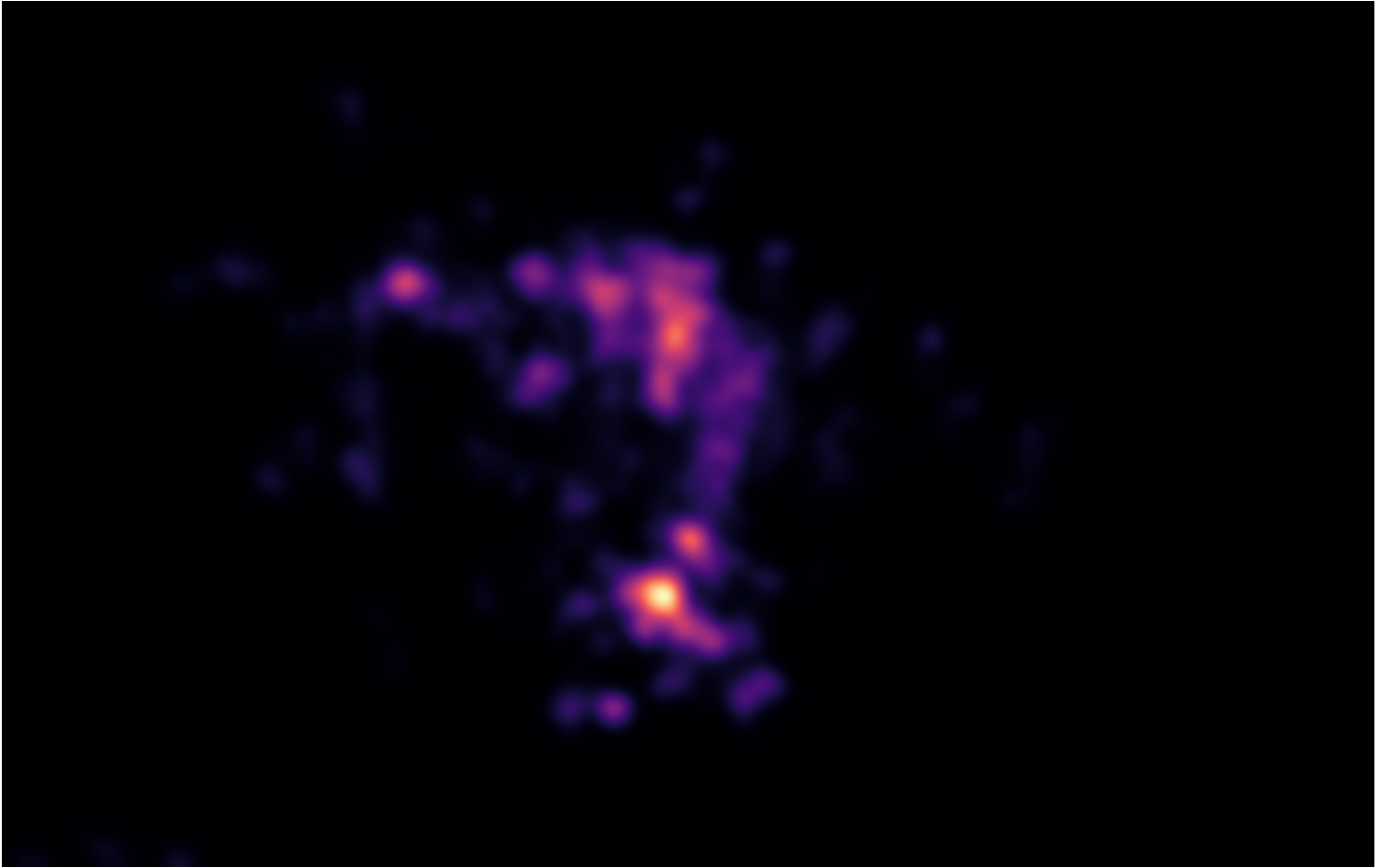}
\end{minipage}\hbox{ }
\begin{minipage}{0.45\textwidth}
\centering
\includegraphics[width=1.0\textwidth]{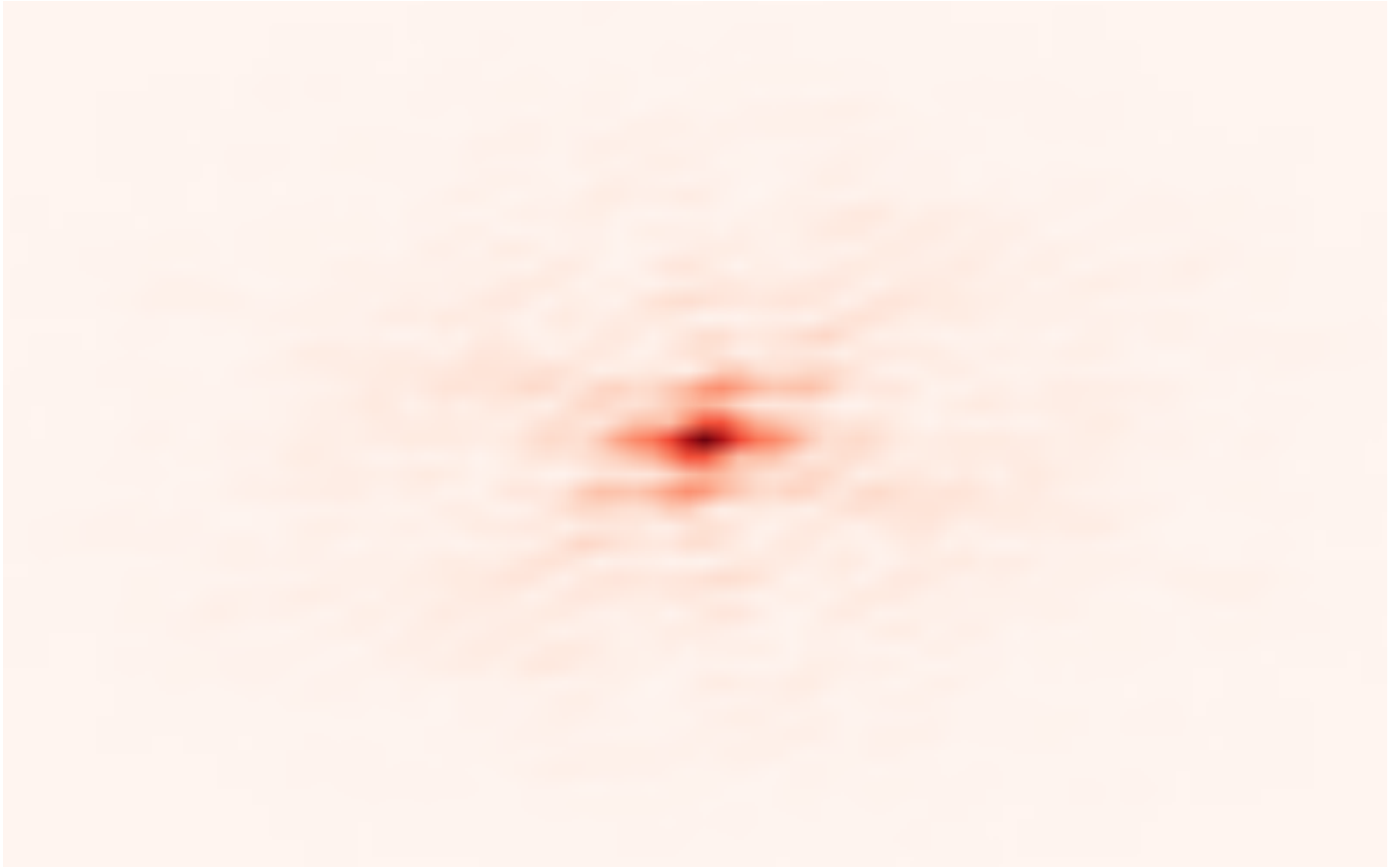}
\end{minipage}
\begin{minipage}{0.45\textwidth}
\centering
\includegraphics[width=1.0\textwidth]{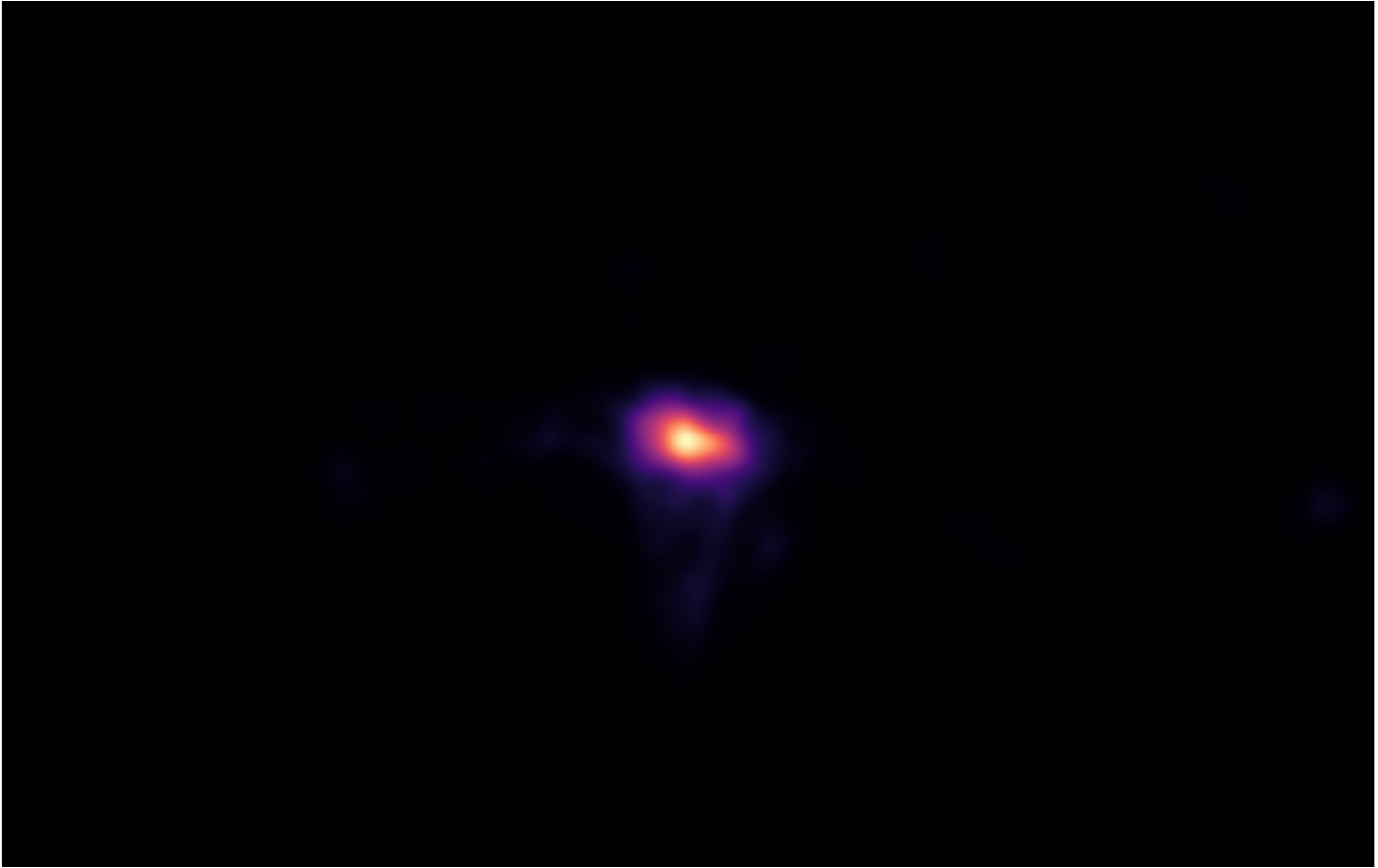}
\end{minipage}\hbox{ }
\begin{minipage}{0.45\textwidth}
\centering
\includegraphics[width=1.0\textwidth]{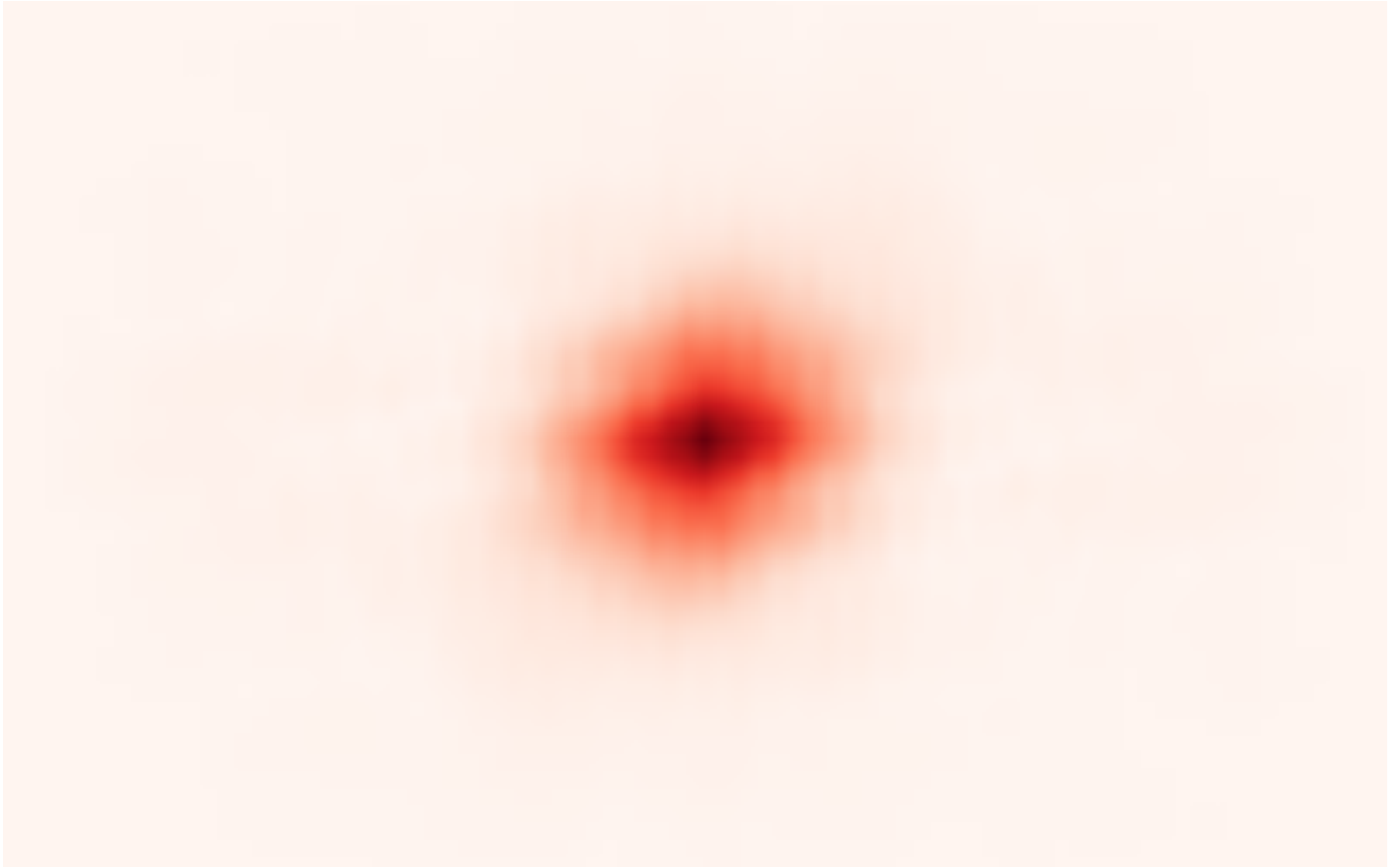}
\end{minipage}
\caption{\textsc{(Left)} Typical FDM recorded during presentation of solidly coloured screens for subjects A \textsc{(top)} and subject B \textsc{(bottom)}.
\textsc{(Right)} DFT'ed FDM corresponding to the ones to the left.}
\label{fig_hm}
\end{figure}

\textbf{Spatial Features -- Fixation Density Map Metrics:}\label{fixation-density-maps}
Examples of normal FDMs from both subjects\footnote{A parallel study with multiple participants, not included in this experiment, indicate that the FDM patterns across subjects are similarly characterized by the amount of central symmetric diffusion, spatial scattering, and vertical/horizontal asymmetrical skewness in their distribution of fixations.} are shown in Fig.~\ref{fig_hm} with their corresponding DFT versions. They are generated from eye traces recorded during the solidly coloured screen epochs and appear very different. The maps shown are somewhat typical for each subject, and although there is variation within-subject, they can generally be classified correctly by a human observer.

Dissimilarity matrices are calculated for the 4 metrics (MSE, 1-Min, KLD and Eucl), for all trials in the first week, for all in the second week and for all in both weeks combined. An example of the best comparison matrices can be seen in Fig.~\ref{fig_blm_all_rclfft_mat}.
From these, ROC and Detection Error Trade-Off (DET) curves are plotted by varying the detection threshold levels, and the accuracy (ACC, proportion of correctly classified outcomes) at the maximum value of F1 (the harmonic mean of precision and recall) is calculated, as is the Area Under Curve (AUC) for the ROC curve and the Equal Error Rate (EER)\footnote{F1, accuracy, area under curve and equal error rates are computed as conventionally.}. The ROC and DET curves corresponding to the matrices in Fig.~\ref{fig_blm_all_rclfft_mat} can be seen in Fig.~\ref{fig_blm_all_rclfft_crv}. All of the computed performance metrics are listed Table \ref{similarity-score-table}. 

\definecolor{lg}{gray}{0.9}

\begin{table}[htb]
\centering
\resizebox{\textwidth}{!}{%
\begin{tabular}{@{}llrrrrrrrrr@{}}
\toprule
 &  & \multicolumn{3}{c}{Week 1} \vline & \multicolumn{3}{c}{Week 2} \vline & \multicolumn{3}{c}{Week 1+2} \\ \midrule
 &  & \multicolumn{1}{l}{ACC} & \multicolumn{1}{l}{AUC} & \multicolumn{1}{l}{EER} & \multicolumn{1}{l}{ACC} & \multicolumn{1}{l}{AUC} & \multicolumn{1}{l}{EER} & \multicolumn{1}{l}{ACC} & \multicolumn{1}{l}{AUC} & \multicolumn{1}{l}{EER} \\
\multirow{4}{*}{FDM} & MSE & 89.0 & 89.7 & 15.5 & 79.5 & 71.5 & 35.3 & 72.2 & 73.6 & 34.5 \\
 & 1-MIN & 85.7 & 92.2 & 15.0 & 87.2 & 94.1 & 16.0 & 79.3 & 85.9 & 21.7 \\
 & KLD & 76.2 & 84.8 & 26.0 & 83.3 & 91.0 & 18.6 & 73.1 & 80.8 & 26.6 \\
 & Eucl & 79.0 & 88.0 & 21.2 & 87.2 & 94.0 & 18.6 & 77.7 & 81.7 & 25.6 \\
\cmidrule(l){2-11}
\multirow{4}{*}{FDM'} & MSE & 79.5 & 89.7 & 21.2 & 97.4 & 99.5 & 3.2 & 74.3 & 82.3 & 27.2 \\
 & 1-MIN & 72.9 & 81.2 & 29.8 & 96.2 & 99.1 & 5.8 & 79.1 & 88.8 & 21.3 \\
 & KLD & 72.9 & 74.9 & 36.0 & 91.0 & 92.9 & 10.9 & 76.1 & 83.0 & 25.9 \\
 & Eucl & 76.2 & 84.0 & 29.3 & 94.9 & 97.2 & 9.6 & 85.6 & 92.0 & 15.8 \\
\cmidrule(l){2-11}
\multirow{4}{*}{DFT} & MSE & 94.3 & 98.6 & 6.4 & 100.0 & 100.0 & \cellcolor{lg} 0.0 & 91.8 & 96.0 & 8.5 \\
 & 1-MIN & 90.5 & 96.8 & 10.7 & 100.0 & 100.0 & \cellcolor{lg} 0.0 & 93.0 & 98.2 & 7.6 \\
 & KLD & 89.5 & 96.4 & 11.2 & 100.0 & 100.0 & \cellcolor{lg} 0.0 & 93.0 & 98.0 & 8.3 \\
 & Eucl & 99.0 & 100.0 & \cellcolor{lg} 1.7 & 100.0 & 100.0 & \cellcolor{lg} 0.0 & 97.7 & 99.6 & \cellcolor{lg} 2.4 \\
\cmidrule(l){2-11}
\multirow{4}{*}{DFT'} & MSE & 97.1 & 99.1 & 5.0 & 100.0 & 100.0 & \cellcolor{lg} 0.0 & 91.4 & 94.9 & 11.1 \\
 & 1-MIN & 95.2 & 98.6 & 6.9 & 100.0 & 100.0 & \cellcolor{lg} 0.0 & 95.5 & 99.0 & \cellcolor{lg} 4.5 \\
 & KLD & 92.9 & 97.8 & 9.7 & 100.0 & 100.0 & \cellcolor{lg} 0.0 & 94.8 & 98.8 & 5.6 \\
 & Eucl & 99.5 & 100.0 & \cellcolor{lg} 0.7 & 100.0 & 100.0 & \cellcolor{lg} 0.0 & 97.7 & 99.6 & \cellcolor{lg} 2.4 \\ \cmidrule(l){2-11} 
\end{tabular}
}
\caption{
\setlength{\parindent}{1em}
Effects of using different dissimilarity scores and domains, showing how well the biometric samples are correctly classified as a match or a non-match, compared to ground truth of mated vs non-mated origins. The Accuracy (\textsc{Acc}) is given at the  threshold value with the highest F1 score. (\textsc{Auc}) is the Area Under Curve and (\textsc{Err}) the Equal Error Rate.
The metrics are based on comparing either original FDMs (\textsc{Fdm}), recalibrated FDMs (\textsc{Fdm'}), DFT'ed FDMs (\textsc{DFT}), or DFT'ed recalibrated FDMs (\textsc{Dft'}), with each dissimilarity score: MSE, 1-Min , KLD , and Eucl.
Trivially identical comparisons are not included in the metrics.
}
\label{similarity-score-table}
\end{table}

As can be seen, the DFT based spatial-frequency approaches generally show a significantly better performance than basing the comparison on the spatial domain of the original FDMs only, and the 1-Min and Eucl metrics consistently\footnote{This holds when looking at optimal threshold values based on, and used within, the same analysed set of of feature vectors.} show the best performance in all four domains. The recalibration has a significantly positive effect on the comparison performance in the spatial domain in week 2, but deteriorates the performance of week 1 spatial domain comparisons\footnote{This could point to the need for an improved re-calibration routine in case there are outliers in the data.}. The effect of the re-calibration is also inconclusive in the spatial-frequency domain, but has a much smaller effect here.

\begin{figure}[htb]
\centering
\begin{minipage}{0.45\textwidth}
\centering
\includegraphics[width=1.0\textwidth]{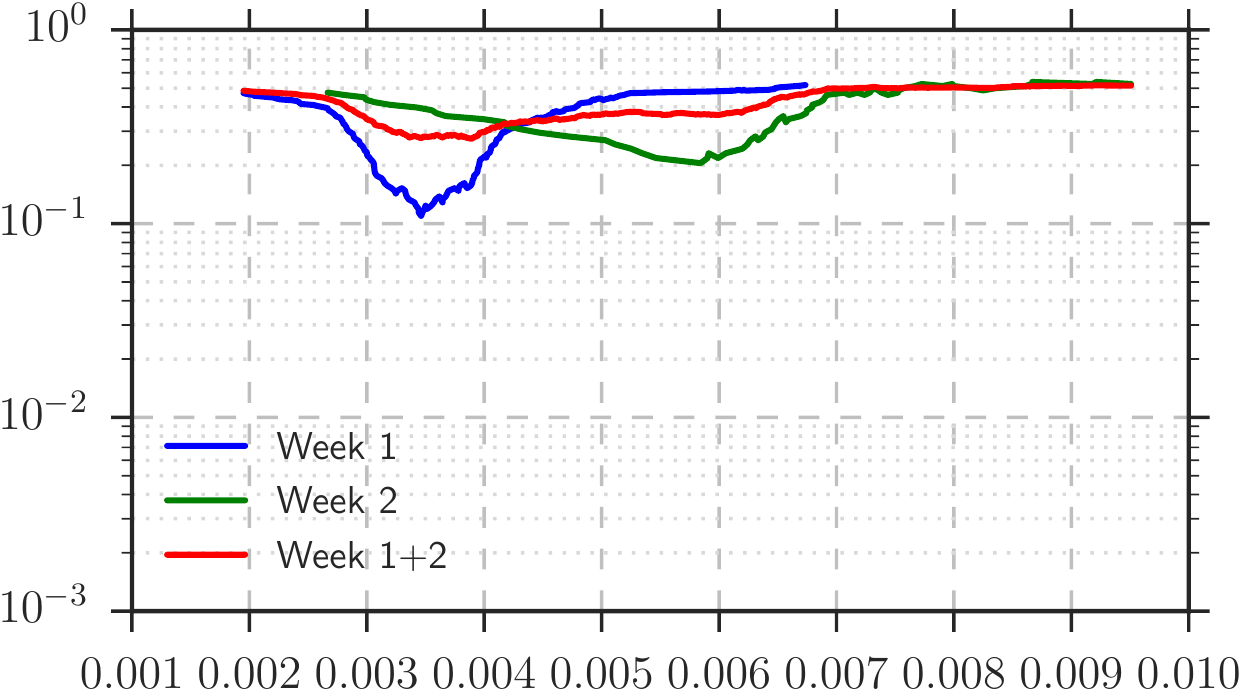}
\end{minipage}\hfill
\begin{minipage}{0.45\textwidth}
\centering
\includegraphics[width=1.0\textwidth]{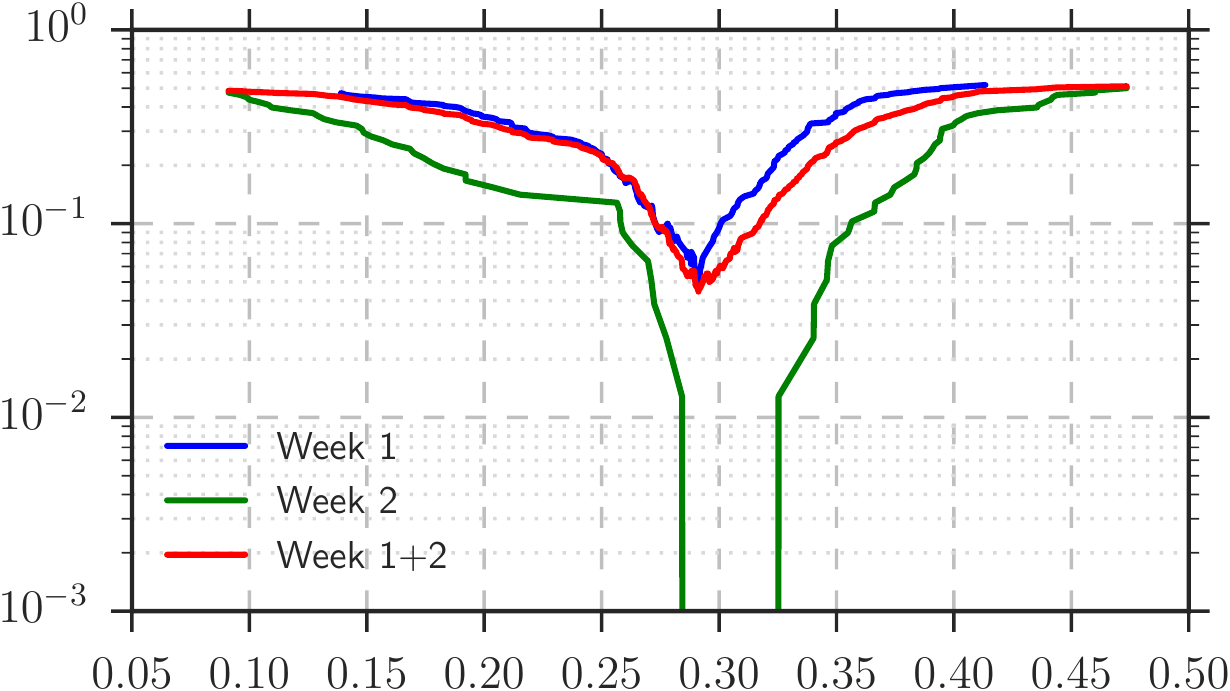}
\end{minipage}
\caption{Examples of total error levels $(1-\operatorname{ACC})$ vs threshold value for (\textsc{Left}) 'misaligned' MSE based comparisons for original DFMs and (\textsc{Right}) 'better aligned' 1-Min based comparison in the DFT domain. Blue, green and red denote week 1, week 2 and weeks 1+2 combined respectively.}
\label{fig-err-level}
\end{figure}

The best performing set of dissimilarity scores show an EER of 0.7\% within week 1 and 0\% (no errors in classification) within week 2; this variation is within the expected statistical variation for a system operating at this level. Between the two weeks, the best result is at 2.4\%. This does \textit{not} however imply that a system trained on one week would automatically give excellent results when tested on the other weeks data. As can be seen in Fig.~\ref{fig-err-level}, the optimal threshold value resulting in lowest error levels for one week neither always coincides with that of the other week nor that of the weeks combined.\footnote{There are differences in the behaviour of the metrics: MSE shows inconsistent alignment of optimal threshold values between weeks in the spatial domain but appear more consistent in the DFT domain, where all metrics generally appear more consistent. This mandates further study before one metric can be recommended in favour of the others.}

The relatively stable and similar performance between and over weeks might give hints on the stability of the FDMs, although the very limited number of subjects makes it impossible to draw any firm conclusions.

Other features extracted from the FDMs have also been tested, such as the spatial variance and skewness of the distributions (for instance, subject A has a tendency to have more fixations to the left side of the screen), but even though there is a clear clustering of data values for each subject they apparently do not provide better comparison scores, and also hold lower entropy, and hence will not be discussed further here.

\begin{figure}[htb]
\centering
\includegraphics[width=1.0\textwidth]{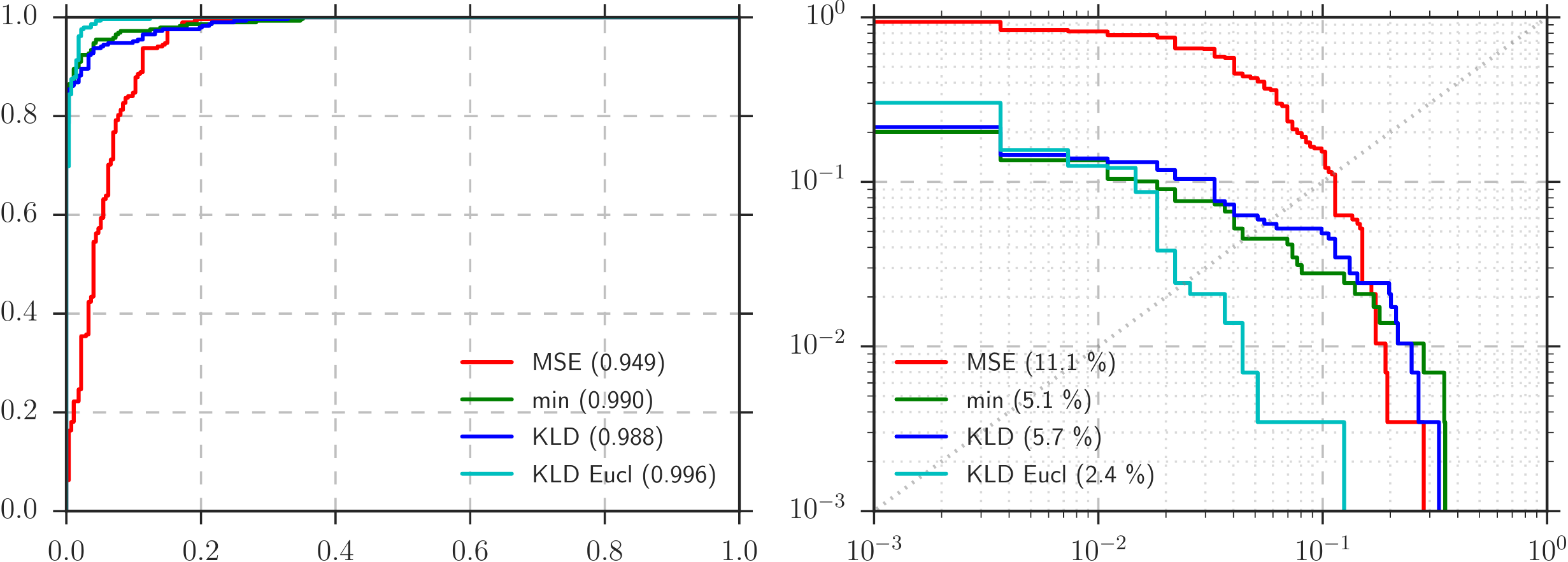}
\caption{
\setlength{\parindent}{1em}
Best performing \textsc{Roc} and \textsc{Det} curves, corresponding to Fig.~\ref{fig_blm_all_rclfft_mat}, with dissimilarity scores: MSE (\textsc{Red}), 1-Min (\textsc{Green}), KLD (\textsc{Blue}), and Eucl (\textsc{Cyan}) over DFT'ed FDMs across all trials. The \textsc{Roc} curve (\textsc{Left}) plots the \textit{True Positives} (y-axis) vs the \textit{False Positives} with resulting \textsc{Auc} of 94.9\%, 99.9\%, 98.8\% and 99.6\% respectively.
The \textsc{Det} curve (\textsc{Right}) plots the \textit{False Negatives} (y axis) vs the \textit{False Positives} on logarithmic scales. The intersection with the line from origo to $(1,1)$ is the point of Equal Error Rate (\textsc{Eer}), of 11.1\%, 4.5\%, 5.6\% and 2.4\% respectively. See Tab. \ref{similarity-score-table}.
}
\label{fig_blm_all_rclfft_crv}
\end{figure}

\section{Conclusion}\label{conclusion}

It has been demonstrated that using a DFT FDM spatial-frequency domain based approach in some instances can give improved performance to biometric recognition systems based on dissimilarity scores, compared to earlier proposed gaussian-filtered FDM spatial domain dissimilarity scores. Performance over the three month period achieved an EER of 2.4\% overall and 0.7\%/0.0\% within each individual week in the best possible case, with corresponding AUC scores of 99.6\% and 100.0\% within weeks.

No significant differences between weeks, time of day, office location or environmental condition was noted, hinting at the stability of the signatures within each individual even over extended periods.

Differences in simple time-to-target values between subjects were also demonstrated, as were directionally dependent differences, and although they also appeared stable over time and conditions, they did not allow statistically significant independent recognition of the subjects. Hence, in a setup similar to the present, with low resolution eye tracking equipment in varying environmental conditions, a temporal-spatial approach deploying the full OPM might be required; simple time-to-target methods might initially be more useful as part of liveness detection.

However, as the top-down and bottom-up modes complement each other, inherently reflecting different human behavioural systems, combining dissimilarity scores based on both would suggest itself as an enhancement compared to treating the two domains independently and should receive continued research efforts. 
As a means to solicit suitable eye trace responses, a combination of salient stimuli with a known timing and non-salient stimuli, even solidly coloured screens, appears within reason and should also be explored further.

\begin{figure}[htb]
\centering
\includegraphics[width=0.80\textwidth]{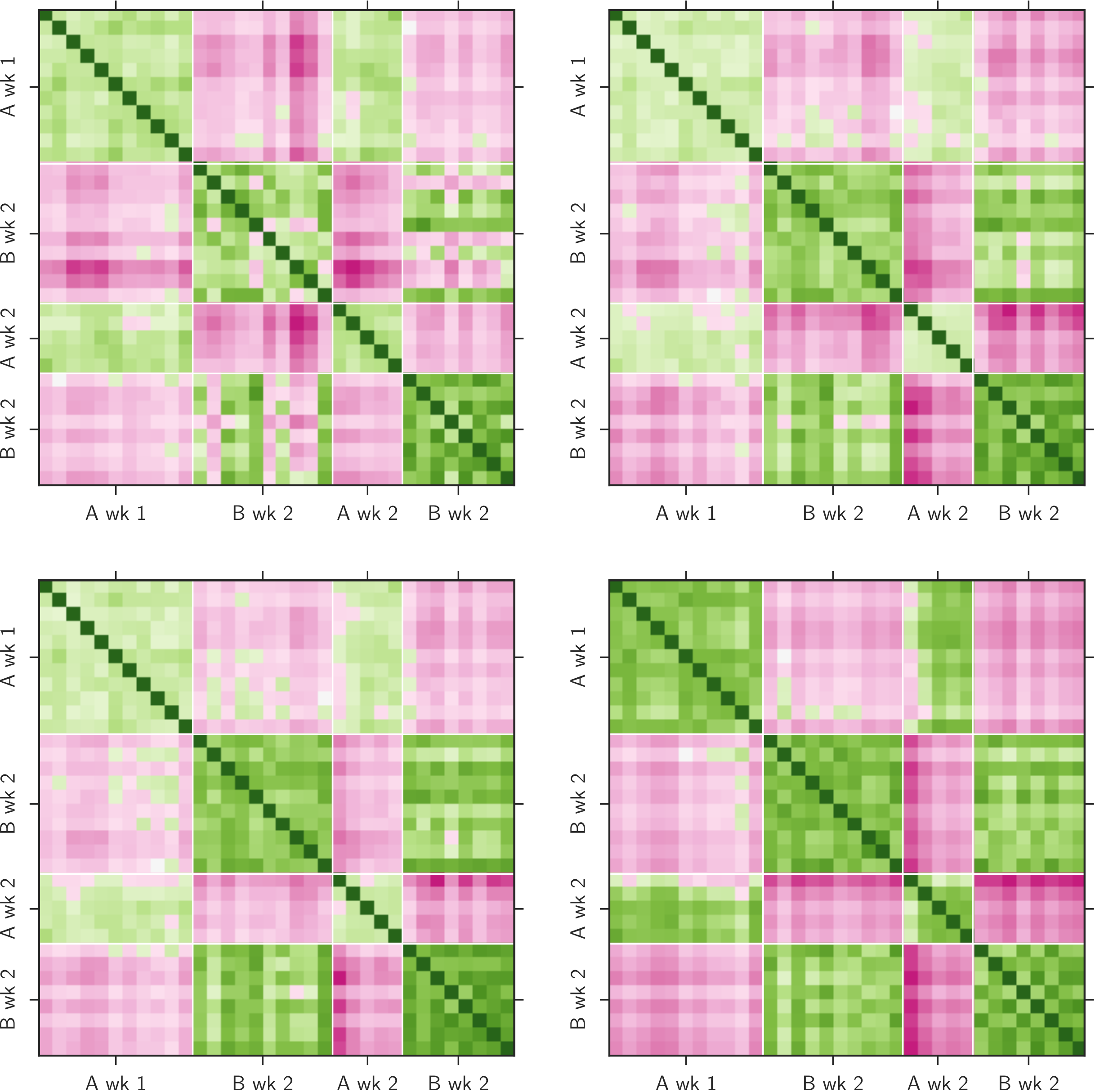}
\caption{
\setlength{\parindent}{1em}
Dissimilarity Matrices based on scores: MSE (\textsc{Top Left}), 1-Min (\textsc{Top Right}), KLD (\textsc{Bottom Left}), and Eucl (\textsc{Bottom Right}) over all DFT'ed FDMs.
Each matrix is divided into 4x4 groups. The first two groups (\textsc{Top} and \textsc{Left}) represents week 1, and the last two groups (\textsc{Bottom} and \textsc{Right}) week 2. Within weeks, the first subgroup is subject A and the last subject B. Each row/column represent a single trial.
Stronger red indicates higher scores (decreasing similarity); darker greens indicate a lower score. The threshold between red and green corresponds to the highest f1 score achievable.
See Fig.~\ref{fig_blm_all_rclfft_crv} for corresponding ROC and DET curves.
}
\label{fig_blm_all_rclfft_mat}
\end{figure}

\bibliography{Library}

\end{document}